\documentclass[prc,preprint,showpacs,endfloats]{revtex4}
\usepackage{graphicx}
\usepackage{bm}
\usepackage{float}
\usepackage{amsmath}
\usepackage{epstopdf}
\usepackage{caption}

\begin{document}


\title{Kinetic energy in the collective quadrupole Hamiltonian from the experimental data}

\author{R.V. \surname{Jolos}}
\author{E.A. \surname{Kolganova}}
\affiliation{Joint Institute for Nuclear Research, 141980 Dubna, Russia}
\affiliation{Dubna State University, 141980 Dubna, Russia}

\begin{abstract}
Dependence of the kinetic energy term of the collective nuclear Hamiltonian on collective momentum is considered.
It is shown that the fourth order in collective momentum term of the collective quadrupole Hamiltonian generates a sizable effect on the excitation energies and  the matrix elements of the quadrupole moment operator. It is demonstrated that the results of calculation are sensitive to the values of some matrix elements of the quadrupole moment. It stresses the importance
for a concrete nucleus to have the experimental data for the reduced matrix elements of the quadrupole moment operator taken between all low lying states with the angular momenta not exceeding 4.
\end{abstract}

\maketitle

The Hamiltonian of the collective nuclear model introduced by A.Bohr~\cite{Bohr52} is a sum of the kinetic  and the potential energy terms. Significant progress has been achieved in understanding of the functional dependence of the potential energy on the collective coordinates. There are several methods of calculation of the potential energy based on the microscopic nuclear models. The potential energy plays a principal role in description of nuclear shape evolution, i.e. transition from spherical to deformed shapes. Generally, the potential energy is a complicated function of two invariants constructed using the collective quadrupole coordinates $\alpha_{2\mu}$. Namely, $(\alpha_2, \alpha_2)_0$ and $(\alpha_2, \alpha_2, \alpha_2)_0$.

At the same time, the kinetic energy term plays as important role in description of collective nuclear dynamics as the potential energy. However, much less progress has been achieved in understanding of a role of different terms which can be presented in the kinetic energy. As for expression for kinetic energy its important ingredient is a mass tensor. It was known from the general expressions obtained within the Generator coordinate method~\cite{Wheeler53,Wheeler57,RingSchuck} and Adiabatic Time Dependent Hartree Fock method ~\cite{Vautherin,Flocard,Villars} that the mass tensor has a complicated dependence on collective coordinates. Nevertheless, in practice, it was often assumed that the mass coefficient can be considered as a constant in the analysis of properties of the low-lying excited states. We should mention, however, that a deformation dependent mass tensor has been considered in~\cite{BM2,Greiner1,Greiner2,Gneuss,Kumar,Petkov}.
Only in description of nuclear fission  where parameters of a nuclear shape undergo considerable variations, dependence of mass coefficient on the deformation parameter was taken into account from the beginning (see \cite{fission2016} and refs. therein).
Some years ago it was shown in \cite{RV2007,RV2009} based on the experimental data for the
excitation energies of the low-lying nuclear excited states and E2 transition probabilities and used in the theoretical analysis \cite{Bonatsos2013,Bonatsos2015,Fraser2011,YAF2013} that the mass parameter in the Bohr Hamiltonian is, in fact, a tensor depending on the shape variables.

The other important feature of the kinetic energy term is its dependence on the collective momentum. It is usually assumed \cite{RingSchuck,Vretenar}
 that it is possible to be limited only by the terms quadratic in collective momentum. However, using the Generator coordinate method~\cite{Wheeler53,Wheeler57,RingSchuck} or the Generalized density matrix method~\cite{Belyaev1,Belyaev2,Belyaev3,Belyaev4}
 it is possible to show that, generally, the expression of the collective Hamiltonian contains all degrees of the square of the collective momentum. Usually terms of the order higher than the square assuming adiabaticity of the collective motion with respect to the single particle one are neglected. Nevertheless, it is interesting to obtain some information on the importance of the neglected terms basing on the experimental data.
 
 In order to stress ones more the important role of the potential energy we should mention that a satisfactory description of the collective features of the spherical, transitional and deformed nuclei has been obtained with the potential energy taken as a complicated function of the collective coordinates (in some cases with several minima) but with a simplest form of the kinetic energy. The simplest form means that only a quadratic in collective momentum term with a constant mass coefficient is taken into account. Therefore, realizing that potential energy is a very complicated function of collective coordinates, and aiming to obtain information about that part of the Hamiltonian that depends on the collective moment
 we need the relations independent on the potential energy. As it is shown below these relations can be obtained by considering the ground state averages of a double and a fourth order commutators of the collective Hamiltonian with the quadrupole moment operator. These expressions do not contain the potential energy. The aim of the present paper is to estimate the effect of the fourth order in the collective momentum term in the Hamiltonian. To our knowledge this problem was not yet analyzed in the literature.

The method which is used below to achieve the aim of the paper is a continuation of the method applied in \cite{RV2007,RV2009} to derive energy weighted sum rules for the E2 transition probabilities. In the Generalized Nuclear Model of A.~Bohr and B.~Mottelson \cite{BM2} the operator  of the electric quadrupole moment $Q_{2\mu}$ is a function of the collective coordinates $\alpha_{2\mu}$ only. For this reason double commutator $[[H,Q_{2\mu}],Q_{2\mu'}]$ does not depend on the potential energy and is proportional to the inverse mass tensor. Taking into account the matrix elements of the double commutator we obtain an expression for it in terms of the energies of the collective states and the reduced matrix elements of the quadrupole moment operator. The values of these quantities can be extracted from the experimental data. We can also calculate the fourth order commutator which in the simplest case can be taken as
\begin{eqnarray}
\label{S0}
{\hat S}=\sum_{\mu,\nu}(-1)^{\mu+\nu}[[[H,Q_{2\mu}],Q_{2-\mu}],Q_{2\nu}],Q_{2-\nu}].
\end{eqnarray}
If the kinetic energy contains only term quadratic in collective momentum the last commutator is equal identically to zero. Thus, taking into account the matrix elements of the fourth order commutator of the Hamiltonian with the operator of the quadrupole moment and expressing it through  experimental data we obtain information on that term of collective Hamiltonian which contains the operator of the collective momentum in the fourth order. For shortness, we denote this term below as $\hat T_4$.

Using the full set of intermediate states we obtain the following expressions for the ground state average of ${\hat S}$
\begin{eqnarray}
\label{S1}
S \equiv \left<0_1|\hat{S}|0_1\right> & = & - 8 \sum_{n,m,k} E(2^{+}_n)\left<0_1||Q||2_k\right>\left<2_k||Q||0_m\right>\left<0_m||Q||2_n\right>\left<2_n||Q||0_1\right>  \nonumber \\
 &+& \frac{4}{5}\sum_{I,n,k,k'} (-1)^{I} E(I_n) \left<I_n||Q||2_k\right>\left<2_k||Q||0_1\right>\left<0_1||Q||2_k'\right>\left<2_k'||Q||I_n\right>  \\
  &+& 2 \sum_{n,k,k'} E(0_n) \left<0_1||Q||2_k\right>\left<2_k||Q||0_n\right>\left<0_n||Q||2_k'\right>\left<2_k'||Q||0_1\right>. \nonumber \nonumber
\end{eqnarray}

This expression contains the sums over the eigenstates of the Hamiltonian with angular momenta $I=0,2,3,4$. It is clear that in order to perform calculations of $S$ for concrete nuclei we have to restrict the summations in (\ref{S1}) by the limited number of collective states because of the absence of the experimental data. The matrix elements of $Q_2$ between the low-lying and high-lying collective states decrease quickly enough with differences in excitation energies. This fact gives us some grounds for restriction of the sums in Eq.(\ref{S1}) by a small number of terms. However, these decreasing matrix elements are multiplied by the excitation energies. Thus, question of convergence of the sum in (\ref{S0}) is open.

Below we restrict summation in (\ref{S1}) by the following collective positive parity states: $0^{+}_{1}$, $0^{+}_{2}$, $2^{+}_{1}$, $2^{+}_{2}$, $2^{+}_{3}$ and $4^{+}_{1}$. Only these states contribute into $S$ in the spherical and rotor limits of the Bohr Hamiltonian.

In the limit of the spherical harmonic oscillator of the Bohr-Mottelson model only the following matrix elements are not equal to zero. In the  units of $\left<0_1||Q_2||2_1\right>$ they are
\begin{eqnarray}
\left<0_2||Q_2||2_1\right>/\left<2_1||Q_2||0_1\right>=\sqrt{\frac{2}{5}},\quad
\left<2_2||Q_2||2_1\right>/\left<2_1||Q_2||0_1\right>=\sqrt{2}, \quad  \nonumber \\ \left<4_1||Q_2||2_1\right>/\left<2_1||Q_2||0_1\right>=\sqrt{\frac{18}{5}}.\qquad\qquad\qquad\nonumber
\end{eqnarray}
In the rigid rotor limit of this model only the following nonzero matrix elements of $Q_2$ contribute to the expression for $S$:
\begin{eqnarray}
\left<2_1||Q_2||2_1\right>/\left<2_1||Q_2||0_1\right>=-\sqrt{\frac{10}{7}} \quad {\rm and} \quad \left<4_1||Q_2||2_1\right>/\left<2_1||Q_2||0_1\right>=\sqrt{\frac{18}{7}}. \nonumber
\end{eqnarray}
In both cases after substitution of these matrix elements into S we obtain that $S=0$. This is an expected result since Bohr Hamiltonian contains only quadratic in collective momentum terms.

It is shown below that the values of $S$ obtained for some nuclei for which there is a relatively sufficient set of the experimental data and in the dynamical symmetry limits of IBM are not equal to zero. In order to be able to estimate the effect of $\hat T_4$  we derive below some useful expressions. First of all, let us assume that the kinetic energy term $\hat T$ of the collective Hamiltonian has the form
\begin{eqnarray}
\label{ke}
{\hat T}= \frac{1}{2B}\sum_\mu \pi^+_{2\mu}\pi_{2\mu} + D \left( \sum_\mu \pi^+_{2\mu}\pi_{2\mu} \right)^2,
\end{eqnarray}
where $\pi_{2\mu}=-i\hbar\frac{\partial}{\partial \alpha_{2\mu}}$ and $\alpha_{2\mu}$ is the collective coordinate which is proportional to the quadrupole moment operator in the Bohr-Mottelson model
\begin{eqnarray}
\label{Qq}
Q_{2\mu} = q\alpha_{2\mu}, \qquad q=\frac{3}{4\pi}e Z R_0^2.
\end{eqnarray}
Thus, in our case $\hat T_4= D (\sum_\mu \pi^+_{2\mu}\pi_{2\mu})^2$.

Our task is to estimate the coefficient $D$. To obtain a dimensionless relation let us find an average value of $\hat{T}$ over the ground state
\begin{eqnarray}
\label{ake}
\left<0_1|\hat{T}|0_1\right> = \frac{1}{2B} \left<0_1|\sum_\mu\pi^+_{2\mu}\pi_{2\mu}|0_1\right>
\left( 1 + 2B D
\frac{\left<0_1|(\sum_\mu \pi^+_{2\mu}\pi_{2\mu})^2|0_1\right>}{\left<0_1|\sum_\mu\pi^+_{2\mu}\pi_{2\mu}|0_1\right>}\right).
\end{eqnarray}
An effect of $\hat T_4$  is characterized by the value of quantity
\begin{eqnarray}
\label{XX}
X\equiv
2B D
\frac{\left<0_1|(\sum_\mu \pi^+_{2\mu}\pi_{2\mu})^2|0_1\right>}{\left<0_1|\sum_\mu\pi^+_{2\mu}\pi_{2\mu}|0_1\right>}
\ge 2B D \left<0_1|\sum_\mu\pi^+_{2\mu}\pi_{2\mu}|0_1\right> .
\end{eqnarray}

To express $\left<0_1|\sum_\mu\pi^+_{2\mu}\pi_{2\mu}|0_1\right>$ through $\left<0_1|\sum_\mu Q^+_{2\mu}Q_{2\mu}|0_1\right>$ we apply the procedure which is used in \cite{Blokhintsev} to derive the uncertainty relation.

Consider the positively determined quantity
\begin{eqnarray}
\label{eq6}
 \sum_\mu \left<0_1|(\xi \alpha^+_{2\mu}-\frac{i}{\hbar}(-1)^{\mu}\pi^+_{2-\mu})(\xi \alpha^+_{2\mu}+\frac{i}{\hbar}(-1)^{\mu}\pi^+_{2-\mu})
 |0_1\right> \ge 0,
\end{eqnarray}
where $\xi$ is a real auxiliary variable. This expression can be rewritten as
\begin{eqnarray}
\label{eq7}
J(\xi)= \xi^2 \left<0_1|\sum_\mu\alpha^+_{2\mu}\alpha_{2\mu} |0_1\right> - 5\xi +
\frac{1}{\hbar^2}\left<0_1|\sum_\mu\pi^+_{2-\mu}\pi_{2-\mu}|0_1\right> \ge 0.
\end{eqnarray}

Since this expression is non-negative (for real $\xi$), this means that the roots of $J(\xi)$ are complex.
This is possible only if
\begin{eqnarray}
\label{eq8}
\frac{4}{\hbar^2} \left<0_1|\sum_\mu\alpha^+_{2\mu}\alpha_{2\mu} |0_1\right> \left<0_1|\sum_\mu\pi^+_{2-\mu}\pi_{2-\mu}|0_1\right> \ge 25.
\end{eqnarray}
Using the relation between $Q_{2\mu}$ and $\alpha_{2\mu}$ we obtain
\begin{eqnarray}
\label{eq9}
\left<0_1|\sum_\mu\pi^+_{2-\mu}\pi_{2-\mu}|0_1\right> \ge \frac{25}{4} \frac{\hbar^2}{q^2}{\left<0_1|\sum_\mu Q^+_{2\mu}Q_{2\mu} |0_1\right>},
\end{eqnarray}
where $\left<0_1|\sum_\mu Q^+_{2\mu}Q_{2\mu} |0_1\right> \approx \left<2_1||Q_2||0_1\right>^2$.
Substituting (\ref{eq9}) into (\ref{XX}) we obtain
\begin{eqnarray}
\label{eq10}
X \ge \frac{25}{4} B D \frac{\hbar^2}{q^2}{\left<2_1||Q_2||0_1\right>^2}.
\end{eqnarray}

For further consideration it is convenient to separate in (\ref{S1}) a dimensional factor. Then the quantity $S$ will be presented as
\begin{eqnarray}
\label{eq11}
S = E(2_1) \left<2_1||Q_2||0_1\right>^4 s \left(\frac{E(I_n)}{E(2_1)}, \frac{\left<I'_{m}||Q_2||I_n\right>}{\left<2_1||Q_2||0_1\right>}\right),
\end{eqnarray}
where $s$ depends only on the ratios of the excitation energies and the $E(2)$ reduced transition matrix elements. The last quantities are expressed in terms of the corresponding B(E2)'s with additional information about their signs.

To estimate $X$ we should know the values of the parameters $B$ and $D$. To find both quantities we consider together with $S$ the following quantity
\begin{eqnarray}
\label{B}
t=\left<0_1|\sum_{\mu}(-1)^{\mu}[[H,Q_{2\mu}],Q_{2-\mu}]|0_1\right>.
\end{eqnarray}
By analogy with $S$ this quantity can be expressed through B(E2)'s and the excitation energies of the collective states. With a good accuracy
\begin{eqnarray}
\label{C}
t=-2E(2_1)\left<2_1||Q_2||0_1\right>^2.
\end{eqnarray}
At the same time we can express $t$ and $S$ through the parameters $B$ and $D$ substituting into (\ref{S0}) and ({\ref{B}) the expressions for the kinetic energy (\ref{ke}) and the quadrupole moment operator (\ref{Qq}). As a result we obtain
\begin{eqnarray}
\label{D}
t & = & -\frac{5\hbar^2q^2}{B}\left(1+\frac{28}{5}B D \left<0_1|\sum_\mu \pi^{+}_{\mu}\pi_{\mu}|0_1\right>\right),\\
S & = & 280 \hbar^4 q^4 D,
\label{E}
\end{eqnarray}
where lower boundary of $\left<0_1|\sum_\mu \pi^{+}_{\mu}\pi_{\mu}|0_1\right>$ is given by (\ref{eq9}).
Equating (\ref{D}) to (\ref{C}) and (\ref{E}) to (\ref{S1}) and taking into account that summation in (\ref{S1}) is restricted by the states listed above we obtain the equations for $B$ and $D$. Finally, we get that
\begin{eqnarray}
\label{Bbeta}
D & = & \frac{E(2_1)\left<2_1||Q_2||0_1\right>^4}{280\hbar^4 q^4} s, \\
\frac{1}{B}& = & \frac{2E(2_1)\left<2_1||Q_2||0_1\right>^2}{5\hbar^2 q^2}- 35 D \frac{\hbar^2q^2}{\left<2_1||Q_2||0_1\right>^2},
\label{Dbeta}
\end{eqnarray}
and
\begin{eqnarray}
\label{X}
 &\quad & X \ge \frac{25}{224}\frac{s}{1-5s/16}.
\end{eqnarray}

Equation (\ref{S1}) for $S$ and (\ref{C}) for $t$ contains the reduced matrix elements of $Q_2$. The absolute values of these matrix elements are equal to the square root of the corresponding B(E2)'s. However, signs of these matrix elements generally are not known from the experiment. The exceptions are the quadrupole moments of the first $2^+$ states.

We have determined the signs of $\left<I'_{m}||Q_2||I_n\right>$ calculating them in the rigid rotor and the spherical harmonic oscillator limits. The results obtained in both limits are in a consent. For the intermediate situation the signs of some matrix elements can be determined  using the consistent-Q formalism~\cite{Warner}. There is also known the following relation between the signs of the matrix elements~\cite{Jolos96}
\begin{eqnarray}
\label{eq44}
sign(\left<2_1||Q_2||0_1\right>) = -sign(\left<0_1||Q_2||2_1\right>\left<2_1||Q_2||2_2\right>\left<2_2||Q_2||0_1\right>).
\end{eqnarray}

Before consideration of the concrete nuclei let us calculate the value of $s$ in the dynamical symmetry limits of IBM~\cite{Iachello} for which all necessary reduced matrix elements of $Q_2$ are known. In the $SU(5)$ limit we obtain that $s=-8/5N$. For instance, in the case of a typical spherical nucleus $^{110}$Pd $N=6$ and $s=-0.18$, which corresponds to $X\approx 0.02$. Here and below we mean under $X$ its lower boundary.  In the $SU(3)$ limit $s=-\frac{24}{N(2N+3)}$ and in the case of a typical deformed nucleus $^{156}_{64}$Gd$_{92}$ $N=12$ and $s=-0.07$, which corresponds to $X \approx 0.01$. In the $O(6)$ dynamical symmetry limit $s=8(1-\frac{(N-1)(N+5)}{N(N+4)})$. For $^{196}$Pt $N=6$ and $s=-0.67$, which leads to $X\approx 0.06$. We can see that in the limit of $N \rightarrow \infty$ in all dynamical symmetry limits $S=0$ as it should be because in this limit IBM coinsides with Generalized Nuclear Model.

Let us apply the consideration outlined above to nuclei for which there is a relatively large set of the experimental data. These are so called $X(5)$ nuclei~\cite{Iachello2000,Jolos2009} -- $^{150}$Nd, $^{152}$Sm and $^{154}$Gd. We also consider $^{110}$Pd. There is no an experimental information on the spectroscopic quadrupole moments of the 2$_2$ and 2$_3$ states. Following the rigid rotor model results we have assumed that
\begin{eqnarray}
\label{2224}
\left<2_1||Q_2||2_1\right> = -\left<2_2||Q_2||2_2\right>=\left<2_3||Q_2||2_3\right>. \nonumber
\end{eqnarray}

\begin{table}
\caption{\label{t2} Results of calculations of $s, X$, and $F$ determined by the expressions (\ref{eq11}), (\ref{X}) and (\ref{eqF}) for several nuclei. The experimental data are taken from~\cite{Tonev,Krucken,Svensson,nndc}}
\begin{ruledtabular}
\begin{tabular}{c c c c c  }
& $^{150}$Nd & $^{152}$Sm & $^{154}$Gd  &$^{110}$Pd \\
\hline
s& 0.16 & -1.13 & -0.49  & -0.88 \\
X & 0.018 & 0.09 & 0.047   & 0.077   \\
F & 0.053 & 0.26 & -0.13 & -0.22 \\
\end{tabular}
\end{ruledtabular}
\end{table}
The results obtained  are presented in Table I.
We can see from this table that the correction of ${\hat T_4}$ to the total kinetic energy lies between 2\% and 9\%, i.e. is restricted by 10\%.

It is also interesting to estimate a contribution of ${\hat T}_4$  into the energy weighted sum rule determined by (\ref{B}). This contribution is given by the second term in the circle brackets of (\ref{D}).
We denote it by F
\begin{eqnarray}
\label{eq19}
F \equiv \frac{28}{5} B D \left<0_1|\sum_\mu\pi^+_{2-\mu}\pi_{2-\mu}|0_1\right>.
\end{eqnarray}
Substituting (\ref{eq9}), (\ref{Bbeta}) and (\ref{Dbeta}) into (\ref{eq19}) we obtain
\begin{eqnarray}
\label{eqF}
F=\frac{5}{16}\frac{s}{(1-5s/16)}.
\end{eqnarray}

The results for $F$ are also presented in Table I. The values of $|F|$   are varied in the limits 0.13 -- 0.26. We can see that ${\hat T}_4$ term is more important for calculations of B(E2)'s than for the excitation energies.

It is seen from the Table I that in the case of $^{150}$Nd the value of $s$ is significantly smaller in absolute value than, for instance, in the case of $^{152}$Sm. At the same time the excitation energies of the collective states and the ratios of the reduced $E2$ transition probabilities are, in general, quite similar in all three X(5) nuclei. Careful investigation of sensitivity of $s$ to variation of different E2 reduced matrix elements have shown that the absolute value of $s$ is  sensitive to variations of $\left<2_1||Q_2||2_1\right>$. The experimental value of $Q(2_1)$ for $^{150}$Nd is equal to $-2.0(5)$ e.b.~\cite{Stone} which corresponds to $\left<2_1||Q_2||2_1\right>/\left<2_1||Q_2||0_1\right>$ $= -1.57 \pm 0.4$.
Varying the absolute value of the last ratio from 1.57 to 1.2 we change $s$ from $+0.16$ to $-1.17$. The last value is close to the result obtained for $^{152}$Sm. Variations of some other reduced matrix elements of $Q_2$ are not significant.  This indicates sensitivity of the results of calculations to the value of $\left<2_1||Q_2||2_1\right>$.

In conclusion, it is shown that the fourth order in collective momentum term
of the collective quadrupole Hamiltonian generates a sizable effect on the excitation energies, and especially,
on the matrix elements of the quadrupole moment operator. Its contribution to the collective kinetic energy can achieve 10\%
and to the energy weighted sum rule - 26\%. These estimates are obtained basing on the experimental data on the excitation energies of the collective states and the E2 transitions matrix elements.   It is demonstrated that the results of calculation are sensitive to the values of some matrix elements of the quadrupole moment operator.  It stresses the importance for a concrete nucleus to have the experimental data for the reduced matrix elements of the quadrupole moment operator taken between all low lying states with the angular momenta not exceeding 4.

\bigskip

Authors acknowledge the partial support from the Heisenberg--Landau Program
and the Russian Foundation for Basic Research.

\end{document}